\documentclass[conference]{IEEEtran}
\IEEEoverridecommandlockouts
\usepackage{cite}
\usepackage{amsmath,amssymb,amsfonts}
\usepackage{graphicx}
\usepackage{booktabs}
\usepackage{multirow}
\usepackage{algorithm}
\usepackage{algorithmic}
\usepackage{xcolor}
\usepackage{url}
\usepackage[hidelinks]{hyperref}

\begin{document}

\title{Multi-Spectrogram Test-Time Adaptation for Open-Set Underwater Acoustic Target Recognition on Recording-Ordered Streams}

\author{\IEEEauthorblockN{Jinhang Hu\textsuperscript{1}, Mubai Ding\textsuperscript{2}, and Longyu Jiang\textsuperscript{1}\IEEEauthorrefmark{1}}
\IEEEauthorblockA{\textsuperscript{1}~School of Computer Science and Engineering, Southeast University, Nanjing 211189, China\\
\textsuperscript{2}~School of Information Science and Engineering, Fudan University, Shanghai 200433, China\\
220262413@seu.edu.cn}
\thanks{\IEEEauthorrefmark{1}Corresponding author: Longyu Jiang.}}

\maketitle

\begin{abstract}
In real deployments, underwater acoustic target recognition (UATR) models often face acoustic conditions that differ from those seen during training. These differences can arise from changes in hydrophone frequency responses, propagation channels, and ambient noise. In open-set continual test-time adaptation (TTA), the model must adapt to unlabeled audio while recognizing known ship types and rejecting unseen ship types and background noise. A batch of sixteen consecutive 5-s clips often comes from a single ship. Under this recording order, the in-domain closed-set accuracy of TTA methods that rely on current-batch statistics drops from about 72\% to 42\%. We propose Multi-Spectrogram Test-Time Adaptation (MSTA). MSTA trains two groups of encoders using four spectrogram representations. At test time, the encoders remain frozen, and MSTA computes a score for each clip from concatenated encoder features. The recording-aware encoders are sensitive to background noise, and the jointly trained encoders are sensitive to unseen ship types. MSTA uses the estimated score shift to select an in-domain or cross-domain detector. It adapts the input through spectral equalization, the decision through segment smoothing and an adaptive threshold, and the classifier through an online teacher. MSTA also fuses the predictions of three frozen pretrained audio models (AST, PANNs, and BEATs) solely for ship-type classification. On DeepShip with simulated sea conditions and on a real DeepShip-to-ShipsEar shift, MSTA reaches a mean H-score of 57.2 over three independently generated test streams. On the main stream, it exceeds the strongest improved baseline by 16.6 points. On Paderborn bearing vibration spectrograms, MSTA outperforms the strongest TTA baseline by 15.8 to 25.6 H-score points across three noise conditions.
\end{abstract}


\section{Introduction}
\begin{figure*}[t]
\centering
\includegraphics[width=\textwidth]{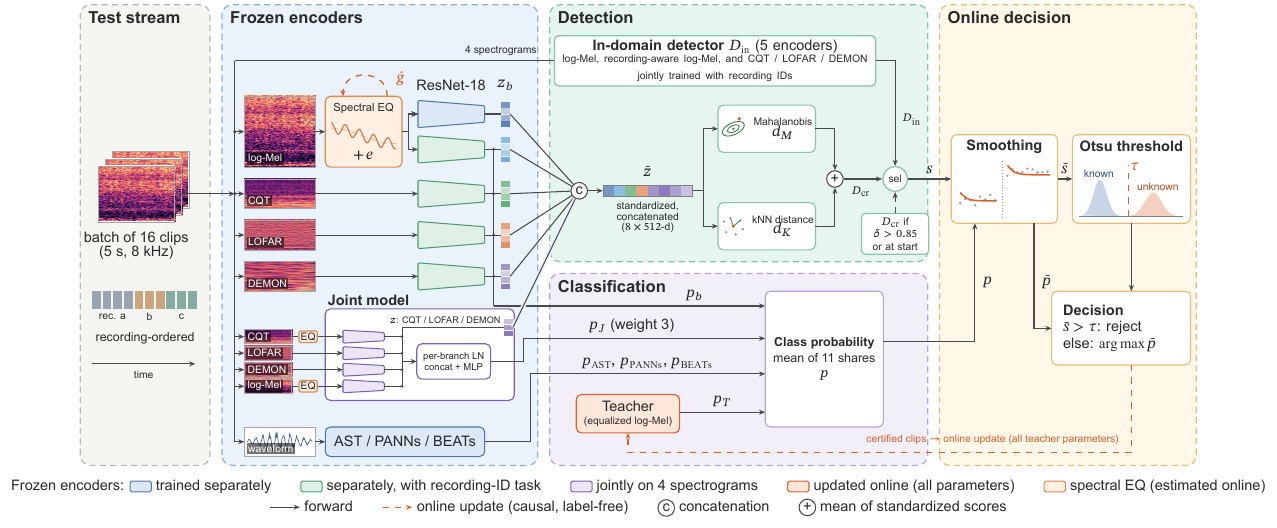}
\caption{MSTA on one batch of 16 clips (the spectrograms are real ones from a cargo-ship clip in the test stream). All encoders are frozen at test time; colors indicate how they were trained. The teacher and the spectral equalizers are updated online. Solid arrows show the forward pass. Dashed arrows show online updates. These updates use only past unlabeled data. The circled c denotes concatenation and $\oplus$ the mean of the two distances after each is standardized on the calibration set. $p_b$ are the class probabilities of the four recording-aware encoders; for this purpose the CQT encoder receives the equalized input.}
\label{fig:overview}
\end{figure*}
Deep models now separate ship types well on public underwater acoustic datasets such as DeepShip~\cite{deepship} and ShipsEar~\cite{shipsear}, but these results usually assume that training and test recordings come from the same conditions. In real deployments, however, the conditions change. A replacement hydrophone has a different frequency response, sea state and water depth change the propagation channel (for example through multipath interference), and the level and spectrum of ambient noise vary over time and place. The incoming audio also contains unknown clips. These clips come from recordings of unseen ship types or from long stretches of pure background noise. A deployed model therefore has to adapt to these shifts without labels while it classifies known ship types and rejects all other clips. This setting is known as open-set continual test-time adaptation (TTA).

TTA~\cite{tent,bnadapt}, continual TTA~\cite{cotta,rmt,gold} and open-set TTA~\cite{ostta,unient,doco} methods were developed on image benchmarks, where every test batch mixes samples of many classes. Underwater audio, however, arrives recording by recording: a passing ship produces dozens of consecutive 5-s clips, so a batch of 16 clips often comes from one recording. Methods that rely on statistics of the current batch, such as batch-normalization (BN) statistics, batch entropy or within-batch partitioning of known and unknown samples, are therefore biased toward that ship. On our in-domain benchmark, TENT and OSTTA with UniEnt+ and open-set GOLD reach 71.6\%--72.1\% closed-set accuracy when clips are shuffled within each scene, but only 41.2\%--42.6\% when clips arrive in recording order, below the 68.8\% of the unadapted source model. Fixed out-of-distribution (OOD) detectors score each clip independently with frozen encoder features~\cite{maha,knn}, so their scores are insensitive to sample order. In our benchmark, these scores offer limited separation between known and unseen ship types, and a fixed calibration threshold cannot track changes in sea conditions.

We propose Multi-Spectrogram Test-Time Adaptation (MSTA, Fig.~\ref{fig:overview}). The same audio can be turned into spectrograms with different physical meanings: log-Mel describes the overall spectral shape, CQT resolves low frequencies finely, LOFAR keeps only line spectra after removing the broadband background, and DEMON describes how the propeller modulates the noise envelope. MSTA trains two groups of encoders on these spectrograms and keeps them frozen at test time. The recording-aware encoders are additionally trained to identify the source recording of each training clip, which makes them sensitive to background noise. The jointly trained encoders process all four spectrograms together, which makes them sensitive to unseen ship types. Their features are concatenated and scored with Mahalanobis and nearest-neighbor distances. At test time MSTA changes only the input (spectral equalization), the decision (segment smoothing, an adaptive threshold and a choice between two detectors according to the estimated shift) and the classifier (an online teacher). The rejection score is computed solely from the frozen encoders and is therefore unaffected by online updates.

Our contributions are as follows.
\begin{itemize}
\item We define an evaluation protocol for open-set continual TTA in UATR, with recording-ordered streams, simulated sea-condition shifts on DeepShip and a real DeepShip-to-ShipsEar shift, and show that methods relying on current-batch statistics fail when clips arrive in contiguous segments from individual recordings;
\item We propose Multi-Spectrogram Test-Time Adaptation (MSTA). To detect unknown clips, MSTA concatenates features from single-spectrogram encoders with those from a model trained jointly on all four spectrogram types. For ship-type classification, MSTA also combines class probabilities from three frozen pretrained audio models (AST, PANNs, and BEATs). At test time, MSTA compares the reference detector's scores on the test stream with its calibration scores to switch between the in-domain and cross-domain detectors. It then uses recent test scores to update the rejection threshold within bounds calibrated for the selected detector.
\item MSTA achieves a mean H-score of 57.2 across three independently generated underwater acoustic test streams. On the main stream, its H-score exceeds that of the strongest improved baseline by 16.6 points. We also apply MSTA to bearing fault diagnosis using vibration spectrograms. On the Paderborn bearing dataset, MSTA outperforms the strongest TTA baseline (our improved version of TENT + UniEnt+) by 15.8–25.6 H-score points across the three noise conditions.
\end{itemize}

\section{Related Work}
\textbf{Test-time adaptation.} TENT~\cite{tent} minimizes prediction entropy on test batches and updates the affine parameters of BN layers, and BN adaptation~\cite{bnadapt} replaces the source BN statistics with test-batch statistics. Continual TTA reduces error accumulation and forgetting under changing shifts with mean teachers~\cite{cotta,rmt}, sample selection~\cite{eata} or a low-dimensional subspace~\cite{gold}. NOTE~\cite{note}, RoTTA~\cite{rotta} and UnMix-TNS~\cite{unmixtns} handle test streams whose labels are temporally correlated, but they assume closed-set data. Open-set TTA methods split each batch into known and unknown samples and minimize or maximize entropy accordingly~\cite{ostta,unient}, filter noisy samples~\cite{sotta}, or use adaptive thresholds and prototypes~\cite{owttt,luostta}; DOCO~\cite{doco} learns a domain-compensation prompt for open-set continual TTA. Previous studies used test-time adaptation to address domain shift in acoustic scene classification~\cite{huang2024acoustictta,liang2024efficientaudio}. In open-set underwater acoustic recognition, clips arrive in recording order, so a batch often comes from a single vessel. This order can reduce the accuracy of methods that rely on current-batch statistics or separate known and unknown samples within each batch.

\textbf{Out-of-distribution detection.} Common post-hoc scores include the maximum softmax probability~\cite{msp}, the energy score~\cite{energy}, ODIN~\cite{odin}, the Mahalanobis distance~\cite{maha} and the $k$-nearest-neighbor distance~\cite{knn}. For several modalities, MultiOOD~\cite{multiood} uses prediction discrepancy between modalities, and Doorenbos et al.~\cite{doorenbos} concatenate normalized features of several modalities before computing a Mahalanobis distance. This is close to our concatenated detector, but their modalities are trained jointly and they do not consider distribution shift or adaptation.

\textbf{Underwater acoustic target recognition.} Common inputs are log-Mel and CQT~\cite{cqt} spectrograms and the LOFAR and DEMON spectra used in sonar processing~\cite{nielsen}; recent work also uses audio models pretrained on AudioSet~\cite{audioset}, such as AST~\cite{ast}, PANNs~\cite{panns} and BEATs~\cite{beats}. Qi et al.~\cite{qi2025timeseries} learn multiscale temporal features directly from acoustic waveforms for underwater target recognition. Kim et al.~\cite{kim26} detect unknown ship types on DeepShip using ODIN and MC dropout~\cite{mcdropout} under the assumption that training and test conditions match. Yuan et al.~\cite{yuan26} study source-free adaptation with target data contaminated by unknown classes and find post-hoc rejection close to random. The UniqueShip dataset~\cite{uniqueship} shows that random splits of DeepShip-like data inflate accuracy by 10--48 points because the same vessel appears in training and test, which matches our analysis in Section~\ref{sec:vessel}. Our spectral equalization follows the idea of cepstral mean normalization~\cite{furui81}, restricted to a basis that describes smooth sensor responses and two-path interference.

\section{Problem Setting and Protocol}\label{sec:setting}
\textbf{Setting.} A source model is trained on labeled source data with known classes $\mathcal{C}$. After deployment, unlabeled test clips arrive in batches $\mathcal{B}_t$ of 16 clips, and the distribution shift changes over time without notice. A clip belongs to a known class, to an unseen ship type (near OOD), or to background noise (far OOD). For each clip the method outputs a known class or ``reject'', using only batch $t$ and earlier batches and no labels.

\begin{figure*}[t]
\centering
\includegraphics[width=\textwidth]{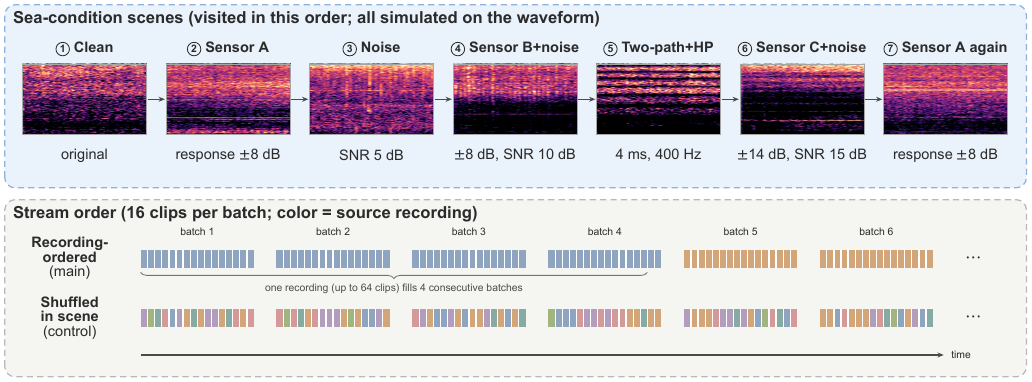}
\caption{In-domain test stream. Top: the seven scenes, each simulated on the waveform (log-Mel thumbnails of one cargo-ship clip, same color scale). Bottom: the two clip orders. In recording order, one recording fills up to four consecutive batches.}
\label{fig:stream}
\end{figure*}

\textbf{In-domain benchmark.} DeepShip contains 609 recordings of four ship types (cargo, passenger ship, tanker, tug). We resample them to 8~kHz, cut 33,770 clips of 5 s, and split the recordings into 426 for training, 90 for calibration and 93 for testing, so that no recording appears in two splits. We use four leave-one-class-out folds, with three known ship types and one unseen type in each fold. Four ShipsEar background-noise recordings provide far-OOD clips. The test stream has 3,769 clips from 97 recordings (at most 64 clips each) and passes through seven scenes (Fig.~\ref{fig:stream}): clean, sensor A, noise at 5~dB SNR, sensor B with noise at 10~dB, two-path interference with a high-pass filter, a strong sensor response C with noise at 15~dB, and finally sensor A again. All shifts are applied to the waveform. A sensor response is a random smooth curve on a logarithmic frequency axis ($\pm$8 or $\pm$14~dB); noise comes from four other ShipsEar background recordings, scaled to the SNR in the 100--2000~Hz band; two-path interference is $y(t)=x(t)+0.6\,x(t-4\,\mathrm{ms})$, followed by a first-order 400~Hz high-pass filter. Because the folds differ, the stream contains 658--1,066 near-OOD clips and 72 far-OOD clips. Clips arrive in recording order (main setting) or are shuffled within each scene (control).

\textbf{Cross-dataset benchmark.} We train on all four DeepShip classes and test on ShipsEar, which differs in waters, equipment and vessels. ShipsEar passenger ships and ocean liners are mapped to passenger ship, RO-RO vessels to cargo and tugs to tug; the other seven vessel types are near OOD and background noise is far OOD. The stream has 2,035 clips from 85 recordings (630 near-OOD and 140 far-OOD clips), and results are averaged over three random recording orders.

\textbf{Test streams and metrics.} To verify that our conclusions do not depend on a particular test stream, we construct two additional streams. For the in-domain part, a different random seed reassigns the scenes, sensor curves and noise segments; for the cross-dataset part, three new recording orders are sampled. The original stream is referred to as the main stream, and all baselines are run on all three streams. The main metric is the H-score~\cite{hscore}, the harmonic mean $H=2AR/(A+R)$ of the known accuracy $A$ (the share of known clips that are accepted and correctly classified) and the rejection rate $R$ (the share of unknown clips that are rejected). We also report the closed-set accuracy (classification of known clips, ignoring rejection), the false rejection rate (FRR) of known clips, and the AUROC between known clips and near- or far-OOD clips. All in-domain results are reported as the average over the four folds, and all cross-dataset results as the average over the three recording orders.

\section{Method}\label{sec:method}
\begin{table*}[t]
\centering
\caption{H-score (\%) of each method on the main stream.}
\label{tab:main}
\setlength{\tabcolsep}{4.5pt}
\begin{tabular}{llccccccc}
\toprule
 & & \multicolumn{2}{c}{In-domain} & \multicolumn{2}{c}{Cross-dataset} & & & \\
\cmidrule(lr){3-4}\cmidrule(lr){5-6}
Type & Method & Ordered & Shuffled & Ordered & Shuffled & Mean & Cross acc. & 3 streams \\
\midrule
\multirow{4}{*}{No adaptation} & Source + MSP~\cite{msp} & 12.2 & 12.2 & 15.8 & 15.8 & 14.0 & 46.9 & 13.8 \\
 & Source + energy~\cite{energy} & 13.8 & 13.8 & 17.0 & 17.0 & 15.4 & 46.9 & 14.9 \\
 & Source + Mahalanobis + residual~\cite{maha} & 26.2 & 26.2 & 30.3 & 30.3 & 28.3 & 46.9 & 28.0 \\
 & Kim et al.~\cite{kim26} & 17.7 & 18.1 & 12.6 & 12.6 & 15.3 & 53.1 & 15.2 \\
\midrule
\multirow{6}{*}{Open-set TTA} & TENT~\cite{tent} + UniEnt+~\cite{unient} & 32.8 & 45.2 & 22.6 & 25.8 & 31.6 & 28.9 & 31.6 \\
 & OSTTA~\cite{ostta} + UniEnt+ & 32.7 & 45.3 & 22.8 & 25.9 & 31.7 & 29.1 & 31.5 \\
 & Open-set GOLD~\cite{gold} & 31.3 & 41.3 & 21.6 & 28.8 & 30.7 & 28.6 & 31.4 \\
 & DOCO~\cite{doco} (AST backbone) & 38.6 & 41.0 & 31.0 & 23.7 & 33.6 & 52.2 & 33.9 \\
 & TENT + UniEnt+, improved$^\dagger$ & 46.5 & 46.3 & 31.0 & 29.4 & 38.3 & 39.3 & 37.5 \\
 & Open-set GOLD, improved$^\dagger$ & 42.6 & 43.7 & 37.2 & 38.4 & 40.5 & 69.5 & 39.9 \\
\midrule
Ours & MSTA & \textbf{60.2} & \textbf{58.5} & \textbf{56.6} & \textbf{53.3} & \textbf{57.1} & \textbf{74.8} & \textbf{57.2} \\
\bottomrule
\end{tabular}
\par\vspace{4pt}
\parbox{\textwidth}{\raggedright\scriptsize Note: In-domain results are averaged over the four DeepShip folds and cross-dataset results over three recording orders, all with a single training seed. Cross acc.: closed-set accuracy on the recording-ordered cross-dataset stream. 3 streams: the mean of the mean H-scores of the three test streams. $^\dagger$Keeps the source BN statistics and separates unknowns across batches.}
\end{table*}

Fig.~\ref{fig:overview} shows the overall structure of MSTA, and Algorithm~\ref{alg:msta} lists the steps for one batch. For each incoming batch, four spectrograms are computed, and log-Mel and CQT are spectrally equalized; the equalized CQT is used for classification only. Frozen encoders then extract features. According to the estimated shift, MSTA picks one of two concatenated detectors, an in-domain one and a cross-domain one, and uses it to compute the unknown score $s$; the class probability $p$ comes from the frozen encoders and the online teacher. Both $s$ and $p$ are smoothed within each recording segment; clips whose smoothed score exceeds the adaptive threshold are rejected, and the others receive the class with the highest smoothed probability. Each batch is decided first, and only then is it used to update the equalizers, the threshold and the teacher.

\begin{algorithm}[t]
\caption{MSTA on one test batch $\mathcal{B}_t$}
\label{alg:msta}
\begin{algorithmic}[1]
\REQUIRE batch $\mathcal{B}_t$; equalizer curves; threshold $\tau_t$; smoothing state; teacher, student and memory
\STATE compute log-Mel, CQT, LOFAR and DEMON; add the current equalizer curves to log-Mel and CQT (CQT for classification only)
\STATE score the batch with the reference detector, update $\delta_t$ (Eq.~\ref{eq:delta}) and choose $D_{\mathrm{in}}$ or $D_{\mathrm{cr}}$
\STATE compute the unknown score $s$ (Eq.~\ref{eq:score}) with the chosen detector and the class probability $p$ (Eq.~\ref{eq:prob})
\STATE detect recording boundaries; smooth $s$ and $\log p$ within each segment to get $\bar s$ and $\bar p$
\STATE reject a clip if $\bar s>\tau_t$, otherwise output $\arg\max_c\bar p_c$
\STATE update the equalizers; certify clips and take one teacher--student step; compute $\tau_{t+1}$ (Eq.~\ref{eq:tau})
\ENSURE one decision per clip
\end{algorithmic}
\end{algorithm}

\subsection{Spectrograms and two groups of encoders}
Each 5-s clip at 8~kHz gives four spectrograms with 157 frames each: log-Mel (256 bands, 100--2000~Hz), CQT (32 bins per octave from 20~Hz, 224 bins), LOFAR (50--1000~Hz, with the broadband background removed frame by frame so that only line spectra remain) and DEMON (the modulation spectrum, 0.5--125~Hz, of the 800--3800~Hz band envelope). Each spectrogram is standardized per clip. All encoders are ResNet-18~\cite{resnet} networks initialized from ImageNet weights (the first-layer kernels of the three color channels are summed) and output 512-dimensional features $z$.

The first group are \emph{recording-aware encoders}, one per spectrogram. The auxiliary recording task is not meant to tell recordings apart at test time; it forces the features to retain recording-level detail, so that background noise, which resembles no training recording, lies far from all of them. During training, a second linear head on $z$ predicts which training recording a clip comes from:
\begin{equation}
\mathcal{L}_{\mathrm{rec}}=\mathrm{CE}_{0.1}(Wz,y)+0.5\,\mathrm{CE}(Az,r),
\end{equation}
where $y$ is the ship-type label, $r$ is the index of the training recording the clip comes from, $W$ and $A$ are the weight matrices of the ship-type head and the recording head, and $\mathrm{CE}_{0.1}$ is the cross-entropy with label smoothing 0.1. The recording task requires telling apart different recordings of the same ship type, which forces the features to keep more detail. Background noise resembles no training recording, so its features lie far from all of them and are easily judged unknown. The second group is a \emph{joint model}. Its four encoders are initialized from separately trained ones; their features are layer-normalized and then concatenated, and a two-layer MLP outputs the fused probability $p_J$. The training loss is
\begin{equation}
\mathcal{L}_{J}=\mathrm{CE}(p_J,y)+\tfrac14\textstyle\sum_{m}\mathrm{CE}(p_m,y),
\end{equation}
where $p_m$ are the outputs of the four branches; during training each spectrogram is zeroed as a whole with probability 0.25. Joint training makes the four spectrograms complement each other, so an unseen ship type that looks plausible in one spectrogram but implausible in another is more easily detected.

\subsection{Concatenated detection}
A detector uses $n$ encoders. Their features $z_1,\dots,z_n$ are standardized per dimension with the mean $\mu_j$ and standard deviation $\sigma_j$ of the source training set and then concatenated:
\begin{equation}
\tilde z=\tfrac{1}{\sqrt n}\big[(z_1-\mu_1)/\sigma_1,\ \dots,\ (z_n-\mu_n)/\sigma_n\big].
\end{equation}
On $\tilde z$ we estimate a mean $m_c$ for each known class and a Ledoit--Wolf shrinkage covariance~\cite{ledoit} $\Sigma$ shared by all classes, which gives the Mahalanobis distance $d_M=\min_c(\tilde z-m_c)^{\top}\Sigma^{-1}(\tilde z-m_c)$. Separately, each encoder's feature is $\ell_2$-normalized, the results are concatenated and normalized again, and the nearest-neighbor distance $d_K$~\cite{knn} is defined as the negative of the 20th-largest cosine similarity between this vector and 8,000 source training clips. Each distance is standardized with the median and interquartile range of its calibration scores, and the two are averaged:
\begin{equation}
s=\tfrac12\big[(d_M-\mathrm{med}_M)/\mathrm{IQR}_M+(d_K-\mathrm{med}_K)/\mathrm{IQR}_K\big].\label{eq:score}
\end{equation}
After concatenation, the shared covariance amplifies combinations that never occur in the source data, for example a clip whose CQT looks like a cargo ship while its LOFAR does not. If one spectrogram alone cannot tell that a clip is unknown, its features occupy only part of the concatenated vector and leave the evidence from the other spectrograms intact. This differs from averaging per-branch scores, whose uninformative branches dilute the high scores given by the other spectrograms.

\subsection{Choosing the detector by the estimated shift}
The two detectors apply the same scoring rule to different combinations of encoder features. On DeepShip recordings under simulated sea conditions, the in-domain detector $D_{\mathrm{in}}$ achieves the higher H-score. When a model trained on DeepShip encounters recordings from ShipsEar, even a familiar ship type can sound different because of the recording equipment or the propagation channel. In this setting, the cross-domain detector $D_{\mathrm{cr}}$ separates unseen ship types from known ones more reliably (Table~\ref{tab:enc}).

Both detectors use the plain and recording-aware log-Mel encoders. They differ in how they represent the other three spectrograms. $D_{\mathrm{in}}$ takes its CQT, LOFAR, and DEMON encoders from a model trained jointly across all four spectrograms for ship-type classification. Each branch also learns to identify the source recording, with a weight of 0.5 on the auxiliary loss. For $D_{\mathrm{cr}}$, each of the three spectrograms has a separately trained recording-aware encoder. Its features are paired with those from the corresponding branch of the joint ship-type classification model.

At test time, MSTA selects the detector according to the estimated shift in the incoming audio. It uses a fixed reference detector to estimate this shift on the same scale before and after a switch. The reference detector uses the same two log-Mel encoders together with plain CQT, LOFAR, and DEMON encoders. We measure the shift relative to source calibration as
\begin{equation}
\delta_t=Q_{0.25}\big(S^{\mathrm{ref}}_t\big)-Q_{0.25}\big(S^{\mathrm{ref}}_{\mathrm{cal}}\big),\label{eq:delta}
\end{equation}
where $Q_{0.25}$ is the 25th percentile, $S^{\mathrm{ref}}_t$ the latest 600 reference scores and $S^{\mathrm{ref}}_{\mathrm{cal}}$ the calibration scores. Most clips in a stream are known, so a rise of this lower quantile as a whole indicates that the whole scene has shifted rather than that unknown targets have appeared. MSTA uses $D_{\mathrm{cr}}$ when $\delta_t>0.85$ and $D_{\mathrm{in}}$ otherwise, and uses $D_{\mathrm{cr}}$ until 150 scores are available. Both the bounds of the adaptive threshold and the certification threshold introduced below are set as quantiles of the calibration scores of whichever detector is currently active.

\subsection{Spectral equalization}\label{sec:eq}
Sensor responses and channels are multiplicative in the linear spectrum, so in the log spectrum they are equivalent to adding a curve to every frame. For each clip we average the raw log-Mel over time to get $q$, update a channel estimate $\hat g$ clip by clip, and derive the compensation curve $e$ from it:
\begin{equation}
\hat g\leftarrow(1-\eta)\,\hat g+\eta\,(q-\bar q_s),\qquad e=c\tanh\big(P(-\hat g)/c\big),
\end{equation}
where $\eta=0.01$, $\bar q_s$ is the mean source spectrum, $P$ is the least-squares projection onto a basis with the constant term removed, and $c$ corresponds to a soft limit of about $\pm$12~dB. The equalized input adds $e$ to every frame and re-standardizes the clip. A sensor response is smooth and is described by eight low-order cosines $\cos(\pi k u)$ ($k=1,\dots,8$, with $u$ the normalized band position). Two-path interference has the frequency response $1+a\,e^{-j2\pi f\Delta}$, whose magnitude forms a comb with period $1/\Delta$ in linear frequency; the eight low-order cosines vary too slowly to fit it. We therefore add the two basis functions $\cos(2\pi f\Delta)$ and $\sin(2\pi f\Delta)$ for each $\Delta=1$, 2, 4, 8~ms, with $f$ the band center frequencies. On the test stream, smooth cosines explain 96\% of the mean spectral deviation caused by sensor responses but only 8\% of that caused by two-path interference; with the comb terms the latter rises to 89\%. CQT is equalized in the same way with smooth cosines only, and the equalized CQT is used only for classification (the class probability of the recording-aware CQT encoder and the input of the joint model).

\subsection{Segment smoothing and adaptive threshold}
Clips of one recording belong to one target. A new segment starts when a clip's feature has a cosine distance above 0.2 to the current segment, or when the root-mean-square standardized deviation of its mean spectrum from the segment exceeds 2. Within a segment the unknown score and the log-probabilities are smoothed exponentially as $\bar s_t=0.8\,\bar s_{t-1}+0.2\,s_t$. If the running rate of new segments exceeds 0.25, the data are evidently not recording-ordered and smoothing is switched off. The threshold is the Otsu split point~\cite{otsu} of the latest 600 smoothed scores $\bar S_t$, clipped to the 70th and 99th percentiles of the calibration scores:
\begin{equation}
\tau_t=\mathrm{clip}\big(\mathrm{Otsu}(\bar S_t),\,q_{70},\,q_{99}\big),\label{eq:tau}
\end{equation}
Before 150 scores are available, $q_{95}$ is used. A clip is rejected if $\bar s>\tau_t$ and is otherwise assigned $\arg\max_c\bar p_c$.

\subsection{Classification and online teacher}
The class probability is an equal-weight average of eleven shares:
\begin{equation}
p=\tfrac{1}{11}\big(p_T+\textstyle\sum_{k=1}^{4}p_k+3\,p_J+p_{\mathrm{AST}}+p_{\mathrm{PANNs}}+p_{\mathrm{BEATs}}\big),\label{eq:prob}
\end{equation}
where $p_k$ come from the four recording-aware encoders, and $p_{\mathrm{AST}}$, $p_{\mathrm{PANNs}}$ and $p_{\mathrm{BEATs}}$ from frozen AST~\cite{ast}, PANNs~\cite{panns} and BEATs~\cite{beats}. For these models, clips are normalized to $-25$~dBFS and resampled to 16~kHz, the embeddings are reduced to 512 principal components, and only a linear classification head is trained per fold. These models classify better across datasets, but their detection scores drift as a whole with the sea conditions, so they are used for classification only. The remaining term $p_T$ comes from an online log-Mel teacher, which is updated as follows. A clip is ``certified'' when its smoothed score is below $q_{70}$, teacher and student agree, and the prediction agrees with the majority vote of its segment; certified clips are stored per class (at most 64 per class). After each batch, the certified clips of the batch plus up to eight clips per class drawn at random from memory are used to update all student parameters with one Adam step (learning rate $3\times10^{-3}$). The loss has two terms: the symmetric cross-entropy between the teacher's prediction on the unaugmented input and the student's prediction on an augmented input (one frequency mask and one time mask), and the KL divergence between the student's and the source model's predictions, which keeps the student from drifting too far from the source model. After each update, the teacher parameters are set to an exponential moving average of the student parameters (coefficient 0.999), and each student parameter is reset to its source value with probability 1\%~\cite{cotta}. The rejection score comes only from frozen encoders, so the teacher affects classification but not detection.

\subsection{Cost}
Per batch, MSTA runs 14 detection encoders (those used by $D_{\mathrm{cr}}$, $D_{\mathrm{in}}$ and the reference detector), the classification branches and one teacher update. On an RTX~3060 laptop GPU this takes about 40~ms per 5-s clip with precomputed pretrained embeddings, and about 25~ms without detector switching.

\section{Experiments}\label{sec:exp}
\subsection{Implementation details and baselines}
Encoders are trained with AdamW for 12 epochs (batch size 64, learning rate $5\times10^{-4}$, weight decay $10^{-4}$) using circular time shifts and frequency and time masking~\cite{specaug} for augmentation, and the epoch with the best calibration accuracy is retained. The joint models are subsequently fine-tuned for another 4 epochs, with a separate set of encoders trained for each fold. The test-time hyperparameters (smoothing rate, equalizer step, threshold window) take the same values in all experiments. The switching threshold 0.85 was set on a tuning stream (in-domain seed 1, cross-dataset order 1); the in-domain part of test stream 1 coincides with the tuning stream, and test stream 2 is held out entirely from tuning.

The baselines fall into two groups. Unadapted methods: the source model (the plain log-Mel encoder) scored with the maximum softmax probability~\cite{msp}, the energy~\cite{energy} or a Mahalanobis distance with a residual term~\cite{maha}, thresholded at the 95th percentile of the calibration scores; and the method of Kim et al.~\cite{kim26} (ODIN~\cite{odin} with MC dropout, on a log-Mel ResNet-18 retrained with dropout). Open-set TTA methods: TENT~\cite{tent} or OSTTA~\cite{ostta} with UniEnt+~\cite{unient}; the open-set extension of GOLD~\cite{gold}; and DOCO~\cite{doco} (with its ViT backbone replaced by AST). For methods that fail on recording-ordered streams we also report improved versions that keep the source BN statistics and separate unknowns using the scores accumulated over the latest 20 batches, instead of splitting only within the current batch. All methods use the same streams, batch size and metrics.

\subsection{Main results}
Table~\ref{tab:main} reports the H-scores on the main stream. The unadapted source models reject few unknowns (17.6\% in-domain with the Mahalanobis score) and reach H-scores of only 12--30. TENT, OSTTA and open-set GOLD reach only 22--33 on recording-ordered streams and recover clearly when clips are shuffled, which shows that the recording order is what breaks them. The improved versions recover part of the gap, but their mean H-score is still at least 16.6 points below that of MSTA on the main stream and at least 17.3 points below over the three streams. DOCO and Kim et al. barely separate known clips from unseen ship types (near-OOD AUROC 47.6--52.0). MSTA is best in all four settings, with a mean H-score of 57.1 on the main stream and 57.6 and 56.8 on the other two streams.

\begin{table}[t]
\centering
\caption{Breakdown on the main stream (\%, in-domain\,/\,cross-dataset).}
\label{tab:break}
\footnotesize
\setlength{\tabcolsep}{1.8pt}
\begin{tabular}{lccccc}
\toprule
Method & Acc. & Rej. & FRR & AUROC$_{\text{near}}$ & AUROC$_{\text{far}}$ \\
\midrule
Source + Maha. & 68.8/46.9 & 17.6/25.5 & \textbf{14.3}/\textbf{20.7} & 55.6/54.1 & 85.3/63.0 \\
TENT + UniEnt+$^\dagger$ & 69.2/39.3 & 53.1/57.4 & 46.5/51.8 & 55.1/55.1 & 79.0/34.5 \\
Open-set GOLD$^\dagger$ & 59.9/69.5 & 53.4/\textbf{59.3} & 47.9/56.9 & 50.0/51.3 & 44.1/42.8 \\
DOCO & 60.7/52.2 & 56.2/58.6 & 54.9/60.1 & 50.7/47.6 & 38.7/18.9 \\
MSTA & \textbf{78.5}/\textbf{74.8} & \textbf{61.2}/57.7 & 32.0/26.5 & \textbf{66.6}/\textbf{68.1} & \textbf{97.8}/\textbf{77.5} \\
\bottomrule
\end{tabular}
\par\vspace{4pt}
\parbox{\columnwidth}{\raggedright\scriptsize Note: All in recording order. Acc.: closed-set accuracy; Rej.: rejection rate of unknown clips; FRR: false rejection rate of known clips. $^\dagger$Improved version.}
\end{table}

In Table~\ref{tab:break}, the adapted baselines reject 46.5\%--60.1\% of known clips and achieve near-OOD AUROCs of 47.6--55.1. MSTA reduces the false rejection rate to 32.0\% in-domain and 26.5\% across datasets and achieves the highest near- and far-OOD AUROCs among the methods shown. The pretrained audio models and the online teacher contribute to the class predictions. The teacher configuration in Table~\ref{tab:abl} (e) raises the recording-ordered cross-dataset closed-set accuracy on the main stream from 68.7\% to 73.4\%.

\subsection{Ablation}\label{sec:abl}
\begin{table}[t]
\centering
\caption{Ablation of the components (mean H-score, \%).}
\label{tab:abl}
\footnotesize
\setlength{\tabcolsep}{2.6pt}
\begin{tabular}{lcccc}
\toprule
Configuration & Main & Stream 1 & Stream 2 & Mean \\
\midrule
(a) Static single log-Mel & 29.6 & 28.3 & 28.8 & 28.9 \\
(b) Static five encoders & 43.8 & 42.1 & 42.1 & 42.7 \\
(c) + Online adaptation & 53.5 & 54.3 & 52.4 & 53.4 \\
(d) Two groups + joint-model fusion & 55.4 & 54.8 & 53.6 & 54.6 \\
(e) Full-network teacher update & 56.0 & 55.5 & 54.2 & 55.2 \\
(f) + Spectral EQ (log-Mel and CQT) & 56.6 & 56.9 & 56.4 & 56.6 \\
(g) + Detector switching (MSTA) & \textbf{57.1} & \textbf{57.6} & \textbf{56.8} & \textbf{57.2} \\
\bottomrule
\end{tabular}
\par\vspace{4pt}
\parbox{\columnwidth}{\raggedright\scriptsize Each stream averages the four settings in Table~\ref{tab:main}. Rows (a) and (b) use fixed calibration thresholds. Spectral equalization starts in (f).}
\end{table}

\begin{table}[t]
\centering
\caption{Results of different detection encoder combinations (\%).}
\label{tab:enc}
\footnotesize
\setlength{\tabcolsep}{2.2pt}
\begin{tabular}{lccc}
\toprule
Detection encoders & Mean H & In-dom. & AUROC near\,/\,far \\
\midrule
5 separate & 55.1 & 58.7 & 64.8\,/\,77.5 \\
Joint-model branches & 53.3 & 56.2 & \textbf{72.4}\,/\,66.2 \\
Recording-aware branches & 54.8 & 57.8 & 62.0\,/\,82.2 \\
Plain + recording-aware (8) & 55.5 & 59.7 & 63.4\,/\,81.6 \\
Joint training + rec.\ IDs ($D_{\mathrm{in}}$) & 53.3 & \textbf{60.8} & 53.3\,/\,\textbf{85.5} \\
Two groups, separately ($D_{\mathrm{cr}}$) & \textbf{56.6} & 58.3 & 68.7\,/\,77.4 \\
\quad with plain CQT/LOFAR/DEMON & 54.7 & 55.7 & 69.2\,/\,76.7 \\
\bottomrule
\end{tabular}
\par\vspace{4pt}
\parbox{\columnwidth}{\raggedright\scriptsize Note: Main stream, with the configuration of Table~\ref{tab:abl} (f) (spectral equalization and full-network teacher updates). All rows use the same classifier configuration and vary the detection encoder set, and every combination includes the plain log-Mel encoder. In-dom.: recording-ordered in-domain H-score; AUROC: recording-ordered cross-dataset stream.}
\end{table}

Table~\ref{tab:abl} starts with two static controls. Both combine Mahalanobis and kNN distances and fix the rejection threshold at the 95th percentile of their source calibration scores. The single log-Mel control (a) uses one encoder for detection and classification. The five-encoder control (b) uses five separately trained encoders for detection and fuses their predictions with those of AST, PANNs, and BEATs for classification. Together these changes raise the three-stream mean H-score from 28.9 to 42.7. Configuration (c) keeps the detection encoders and classifier fusion of (b) and adds segment smoothing, an adaptive Otsu threshold, and teacher updates restricted to the BN affine parameters. The mean H-score rises by 10.7 points to 53.4.

Configuration (d) uses the recording-aware and jointly trained encoders for detection and incorporates the joint model's fused predictions into the classifier. These changes raise the three-stream mean H-score by 1.2 points. The cross-dataset H-score rises from 50.4 to 53.0, while the in-domain H-score moves from 56.4 to 56.2. In (e), the student updates all network parameters and the teacher tracks them by exponential moving average. This configuration raises the mean H-score by 0.6 points. The rejection decisions are identical in (d) and (e). Configuration (f) adds log-Mel and CQT equalization (Section~\ref{sec:eq}). Equalized CQT is used for classification only. The mean H-score rises by 1.4 points, with gains of 0.6, 1.4, and 2.2 across the three streams. In-domain, the average H-score rises by 2.1 points and closed-set accuracy rises from 78.1\% to 82.0\%. Across datasets, the mean H-score rises by 0.7 points. Detector switching in (g) raises the in-domain H-score by 0.5--1.9 points in every stream and order. Cross-dataset H-scores change little. On the main stream, initializing the switch with $D_{\mathrm{in}}$ yields a mean H-score of 56.9, compared with 57.1 when starting with $D_{\mathrm{cr}}$.

Table~\ref{tab:enc} compares encoder combinations for detection. The joint-model encoders are good at near OOD and the recording-aware encoders at far OOD: the former give the highest cross-dataset near-OOD AUROC (72.4) but a far-OOD AUROC of only 66.2, while the latter give 62.0 near-OOD and 82.2 far-OOD. Concatenating the two separately trained groups ($D_{\mathrm{cr}}$) gives near- and far-OOD AUROCs of 68.7 and 77.4 and the highest mean H-score (56.6). The eight-encoder combination of plain and recording-aware branches scores 55.5, compared with 56.6 for $D_{\mathrm{cr}}$ and 55.1 for the five separately trained encoders. Adding the recording task to joint training ($D_{\mathrm{in}}$) gives the best in-domain H-score (60.8) but lowers the cross-dataset near-OOD AUROC to 53.3, so this combination is used only as the in-domain detector.

Several other changes beyond those in the tables brought no gain. Unlike Tables~\ref{tab:abl} and~\ref{tab:enc}, these comparisons use log-Mel equalization together with the injection check (background-noise clips are injected every 400 clips, and the threshold is reset to $q_{95}$ when too many of them are accepted). Using a separate Otsu threshold per class, as in LU-OSTTA~\cite{luostta}, lowered the mean H-score of the five separately trained encoders from 54.0 to 52.1, because unknown clips predicted as one class raise that class's threshold. Giving the online teacher half of the classification weight lowered it from 54.0 to 50.8, since on recording-ordered streams the teacher drifts toward the class of the current ship. Fusing the classifiers with temperatures and weights fitted on calibration data lowered it from 56.0 to 54.2. Adding BEATs to the detector under large shifts, with relaxed threshold bounds, lowered the BN-only version from 55.5 to 53.6: across datasets it raised the far-OOD rejection rate from 68.3\% to 80.2\% but lowered the near-OOD rejection rate from 55.6\% to 45.6\%.

\subsection{Seen and unseen vessels}\label{sec:vessel}
\begin{table}[t]
\centering
\caption{Detection results for test recordings of known ship types, grouped by whether their vessel appears in the training set.}
\label{tab:vessel}
\footnotesize
\setlength{\tabcolsep}{3pt}
\begin{tabular}{lcccc}
\toprule
Test clips & Clips/stream & Rej. & Acc. & AUROC \\
\midrule
Known type, seen vessel & 2120 & 30.0 & 87.9 & 71.0 \\
Known type, unseen vessel & 652 & 45.1 & 65.1 & 54.3 \\
Unseen ship type & 924 & 58.3 & -- & -- \\
\bottomrule
\end{tabular}
\par\vspace{4pt}
\parbox{\columnwidth}{\raggedright\scriptsize Note: Recording-ordered in-domain streams, averaged over the three test streams and four folds. Rej.: share of clips rejected (\%); Acc.: closed-set accuracy (\%); AUROC (\%): against clips of unseen ship types.}
\end{table}

DeepShip is split by recording, but a vessel often has several recordings. After matching recordings to vessel names with the dataset's metadata, 68 of the 92 matched test recordings (74\%) come from vessels that also appear in the training set. Table~\ref{tab:vessel} shows that detection behaves very differently on the two groups: clips of vessels seen in training are rejected 30.0\% of the time, with an AUROC of 71.0 against unseen ship types, whereas clips of known ship types recorded from vessels absent from training are rejected 45.1\% of the time and the AUROC drops to 54.3. A large part of the in-domain ``unknown'' signal is therefore actually ``this vessel was not in training''. This agrees with UniqueShip~\cite{uniqueship} and also explains why the cross-dataset stream, where all vessels are new, remains hard.

\subsection{Bearing vibration data}
To test whether MSTA applies to other data types that arrive segment by segment, we repeat the evaluation on the Paderborn bearing dataset~\cite{paderborn}. The known classes are 6 healthy bearings and 5 bearings with real outer-race damage; 6 bearings with real inner-race damage form the unknown class and appear only in the test stream. Training uses measurements 1--10 of one operating condition and calibration uses measurements 11--13; the test stream passes through four operating conditions in turn (changing the rotational speed, the load torque or the radial force), with 68 recordings and 4,556 clips of 1 s, 35\% of them unknown. In addition, white noise is added to every test clip at 20~dB and 10~dB SNR. Signals are resampled to 12~kHz, the spectrogram frequency ranges are reset for vibration signals, and all encoders are retrained with the same settings; the pretrained audio models and the comb basis are not used.

\begin{table}[t]
\centering
\caption{H-score (\%) on the Paderborn bearing test stream.}
\label{tab:bearing}
\footnotesize
\setlength{\tabcolsep}{4pt}
\begin{tabular}{lccc}
\toprule
Method & Clean & 20 dB & 10 dB \\
\midrule
Source + Mahalanobis + residual & 74.7 & 73.5 & 68.6 \\
Best other baseline & 59.3 & 60.3 & 61.3 \\
TENT + UniEnt+, improved & 53.0 & 53.2 & 57.8 \\
MSTA, unchanged & 65.7 & 66.4 & 56.9 \\
MSTA, shifted upper bound & \textbf{78.6} & \textbf{78.7} & \textbf{73.5} \\
\bottomrule
\end{tabular}
\par\vspace{4pt}
\parbox{\columnwidth}{\raggedright\scriptsize Note: Each number is the mean over the recording-ordered and shuffled orders. Noise is added only to test clips.}
\end{table}

Without any change, MSTA reaches an H-score of only 65.7 without noise, below the 74.7 of the source model (Table~\ref{tab:bearing}): at 900~rpm and at 400~N radial force, the scores of known clips rise as a whole above the 99th percentile of the calibration scores, so the threshold is held at its upper bound and many known clips are rejected. We therefore made one change specific to this dataset: the upper bound becomes $q_{99}+\max(0,\delta_t)$, where $\delta_t$ is the difference between the 25th percentile of the latest 600 scores of the active detector and the 25th percentile of its calibration scores. On the three underwater test streams this change alters the mean H-score by only $-0.1$, so it is not counted as part of the main method. With this change, MSTA exceeds the strongest TTA baseline (our improved version of TENT + UniEnt+) by 25.6 H-score points without added noise, 25.5 at 20~dB, and 15.8 at 10~dB. Relative to the unadapted source model, MSTA gains 3.9--5.2 H-score points. Its AUROC is 89.0\% without noise and 83.1\% at 10~dB (source model: 86.1\% and 79.1\%). Here training and test use the same bearings, and the shift comes only from the operating conditions.

\section{Limitations}
Our evaluation can be extended further. First, the experiments use a single training seed, and results with multiple seeds will be added. Second, the in-domain sea conditions are simulated, and the only real cross-dataset test is from DeepShip to ShipsEar, two datasets that share three known classes, so validation on more real data is needed. Third, detector switching raises the number of detection encoders per batch from 8 to 14 and the run time by about 70\%, which leads to our next step of reducing the computation while keeping the performance.

\section{Conclusion}
We studied open-set continual test-time adaptation for underwater acoustic target recognition and showed that methods relying on current-batch statistics fail when data arrive recording by recording. To address this, we proposed MSTA, which detects unknowns with the concatenated features of two groups of frozen multi-spectrogram encoders and chooses the detector according to the estimated shift. Its adaptation acts only on the input, the decision and the classifier, while the features used for detection stay fixed. Over three test streams MSTA reaches a mean H-score of 57.2, 16.6 points above the strongest improved baseline on the main stream, and, with the adjusted threshold upper bound, it outperforms the strongest TTA baseline on bearing vibration data. This suggests that the method also applies to other domains where data arrive segment by segment.

\bibliographystyle{IEEEtran}
\bibliography{refs}

\end{document}